\documentclass[twocolumn]{svjour3}          
\smartqed  
\usepackage{graphicx}
\usepackage{epstopdf}
\usepackage[utf8]{inputenc}
\usepackage[T1]{fontenc}
\usepackage{amsmath}
\usepackage{gensymb}
\usepackage[misc]{ifsym}
\usepackage[numbers,sort&compress,square]{natbib}
 \journalname{Applied Physics B}
\begin{document}

\title{Optical Frequency Locked Loop for long-term stabilization of broad-line DFB lasers frequency difference
}


\author{Michał Lipka$^1$         \and
       Michał Parniak$^1$	\and
       Wojciech Wasilewski$^1$
}

\authorrunning{Michał Lipka et al.} 

\institute{\Letter\ Michał Lipka \at
              mj.lipka@student.uw.edu.pl
           \and
           $^1$ Institute of Experimental Physics, Faculty of Physics, University of Warsaw, Pasteura 5, 02-093 Warsaw, Poland 
}

\date{Received: date / Accepted: date}

\maketitle

\begin{abstract}
We present an experimental realization of the Optical Frequency Locked Loop (OFLL) applied to long-term frequency difference stabilization of broad-line DFB lasers. The presented design, based on an integrated phase-frequency detector chip, is digitally tunable in real-time, robust against environmental perturbations and compatible with commercially available laser current control modules. We present a simple model and a quick method to optimize the loop for given hardware using only simple measurements in time domain and approximate laser linewidth. We demonstrate frequency stabilization for offsets encompassing entire 4-15 GHz capture range. We achieve $<$ 0.5 Hz long-term stability of the beat note frequency for 1000 s averaging time. Step response of the system as well as phase characteristics closely adhere to the theoretical model.

\end{abstract}

\section{Introduction}
\label{intro}
Lasers frequency difference stabilization is indispensable in multiple modern experimental schemes. Applications range from quantum optics, cold atomic physics and off-resonant light-atom interfaces \cite{Leveque2014,Yim2014,Dabrowski2014,Parniak2016a}, through frequency comb stabilization \cite{Cundiff2003,Jayich2016,Gunton2015,Lezius:16} to precision spectroscopy and sensing \cite{Lyon2014,Matthey2006}.

In multiple applications phase coherence of the two laser fields locked at a frequency offset is not required and a mere frequency lock constitutes a sufficient solution. Nevertheless, one of the most commonly used solution is the Optical Phase Locked Loop (OPLL) \cite{Appel2009,Friederich2010,Wei2016}. In a generic OPLL the Master Laser (ML) and the Slave Laser (SL) are combined and the beat note is measured, compared with a reference value and then the difference is fed through the loop filter and used to tune SL through a fast current modulator. 
However, phase difference has to be kept within tight margins for OPLLs to work, rendering them impractical for broad-line lasers.

Here we present a  simplified version of the OPLL used to stabilize only the long-term ($>100\;\mu \textrm{s}$) frequency drift. Our Optical Frequency Locked Loop (OFLL) setup is based on an integrated phase-frequency detector (PFD) chip that compares the beat note signal of ML and SL with low-frequency reference. Since the output of the detector is proportional to phase difference, we design a single-stage loop filter as a proportional controller with only a small integral term to keep the phase difference in the detector range. The PFD chip may detect very large phase differences and thus can be applied to broad-line laser diodes. The error signal is fed back to a simple current controller with relatively slow response.

Several different methods had been developed for the purpose of frequency locking. These include feedback loops involving Mach-Zehnder interferometer with coaxial cable delay lines \cite{Schunemann1999} or usage of electrical frequency filters \cite{Schilt2008,Strauss2007} to perform frequency to amplitude conversion. These methods suffer from some significant limitations, such as: less compact design, susceptibility to the environmental conditions of the frequency filter or limited tunability. 

We address these issues by employing a method that yields excellent results in phase stabilization \cite{Appel2009} to long-term frequency stabilization of broad-line lasers, like DBF laser diodes applicable in harsh environments \cite{Leveque2014,Lezius:16}. In particular, the compact design is guaranteed by using an integrated PFD chip. Long-term stability is limited only by the frequency reference generator and tuning of the setup may be performed in real-time by reprogramming the PFD chip and the generator.

In the second and the third part of the article we discuss limitations of the OPLL operation and how these are addressed in our solution. The fourth part describes a simple theoretical model we have developed to ease the setup and optimization of the OFLL with generic hardware. The sole process is then described in part five in the form of a step-by-step procedure requiring merely simple time domain measurements.  Part six gives a simple method to modify the model to account for specific hardware characteristics. The last two parts contain the results concerning performance of our realization of the OFLL and the conclusions respectively. 

\section{OPLL limitations}
In a typical OPLL the beat note of ML  and SL is registered on a photodiode detector (PD) and then an electronic mixer is used for comparison of the phase with the reference local oscillator (LO). 
Such construction imposes severe limitations on the maximal phase error  $|\phi| < {\pi}/{2}$ rad due to periodic response of the mixer.
This limitation can be safely neglected in phase-coherence focused applications where $\left<\phi^2\right> \ll 1$ and loop bandwidth exceeds laser linewidth. 
However, in the regime of slowly reacting, broad-line lasers $\left<\phi^2\right>$ can reach thousands. 
This is caused by the intrinsic laser frequency drift within the loop response time.

In feedback loop systems unavoidable delays limit the reaction time of the loop. 
In typically used frequency domain this corresponds to the loop bandwidth $f_u$ --- the maximal frequency at which the loop gain $G(f)$ is above unity. 
The loop bandwidth $f_u$ is bound by the inverse of the loop delay $\Delta \tau$.

The phase fluctuations at frequencies above $f_u$ cannot be compensated by the loop. Therefore, the loop delay is the main reason why the state-of-the-art, high speed electronics is needed to maintain sufficiently high $f_u$ and thus sufficiently low $\left<\phi^2\right>$ for a typical OPLL to function properly.  

\section{Optical Frequency Locked Loop}
If one is concerned merely with long-term frequency drift compensation, slow loop is sufficient, provided it can accommodate relatively large phase fluctuations $\left<\phi^2\right>$. 
This is enabled by replacing a simple mixer with a commercial PFD integrated with programmable dividers in a single chip. 
If the beat note signal is divided by $N$, the maximum phase difference for linear PFD regime becomes $N\pi$.
Allowing for larger phase detours $\left<\phi^2\right>$, the loop bandwidth $f_u$ is no longer required to exceed the laser linewidth. 
Larger acceptable loop delay $\Delta \tau$ follows enabling usage of the commercially available slow laser current controllers in the loop. 
\begin{figure}
 \includegraphics[width=\columnwidth]{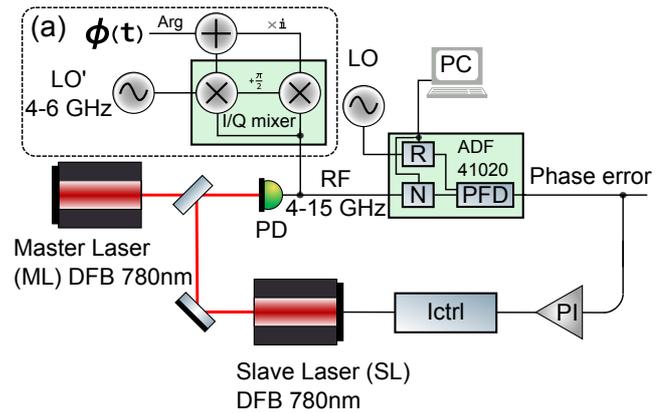}
\caption{OFLL and (a) phase $\phi (t)$ measurement setup. Broad-line (10~MHz) distributed feedback (DFB) Master laser (ML) and Slave laser (SL) frequency difference (RF) is measured as a beat note on a fast photodiode (PD) and fed to a programmable ADF41020 phase-frequency detector (PFD) where its frequency is divided (N) and compared with a frequency divided (R) reference local oscillator (LO). Phase error signal is fed through a proportional-integral controller (PI) to slow laser current controller (Ictrl) closing the feedback loop. In (a) configuration RF signal is fed to In-phase quadrature (I/Q) mixer along with high frequency local oscillator (LO'). The two local oscillators (LO and LO') share a common 10~MHz clock reference. Relative phase of I/Q outputs comprises the RF to LO' phase difference $\phi (t)$.
}
\label{fig:setup}      
\end{figure}

\begin{figure}
 \includegraphics[scale=0.7]{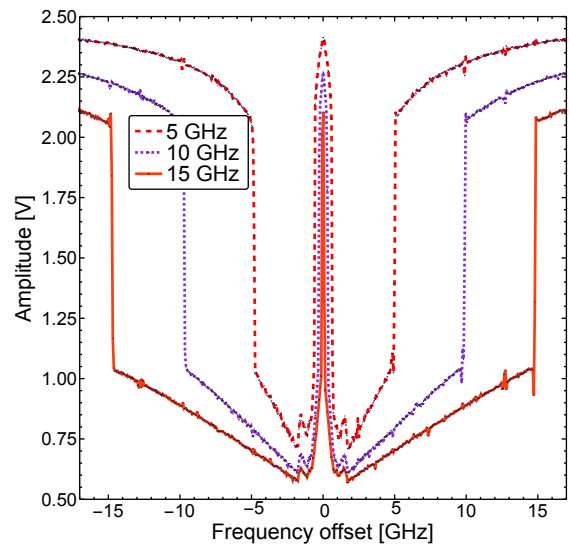}
\caption{ADF41020 phase-frequency detector (PFD) output signal for 3 different locking points during broad $\pm 17\:\textrm{GHz}$ laser frequency scan in OFLL configuration with the loop opened. Note an extreme span of unambiguous error signal. Sharp slopes in the PFD output indicate that up to $15\:\textrm{GHz}$ beat note signals (RF) can be stabilized in this OFLL configuration. }
\label{fig:pfd}       
\end{figure}
In the experiment we use two distributed feedback (DFB) lasers (Toptica DL100). The Master laser (ML) is free running during all measurements. The Slave laser (SL) is controlled by Toptica current controller (DCC110). The PI controller is built using merely one LM358N operational amplifier with several passive components. 
About 100~$\mu$W power from each laser is combined with a fiber coupler (Thorlabs FC780-50B) and the resulting beat note is gathered on a $\sim$10 GHz photodiode (PD) (Finisar HFD6380-418) as depicted in Fig. \ref{fig:setup}. The signal from complementary microwave outputs of the photodiode is fed via an roughly 7~mm long coplanar lines realized on standard FR4 laminate (0.15~mm clearance between 1~mm RF trace and surrounding ground, 1.6~mm thick laminate, ground stitched to continuous plane on the bottom of a board with 0.3~mm vias spaced 0.6~mm apart) to PFD chip (ADF41020) and SMA diagnostic output. 
The photodiode saturates at about 500~$\mu$W total power, providing approximately -10~dBm RF power at 6~GHz.
In this arrangement the PFD operates reliably between 4 and 15 GHz as depicted in Fig.~\ref{fig:pfd}

\section{Simplified loop model}\label{sec:smodel}

\begin{figure}
 \includegraphics[width=\columnwidth]{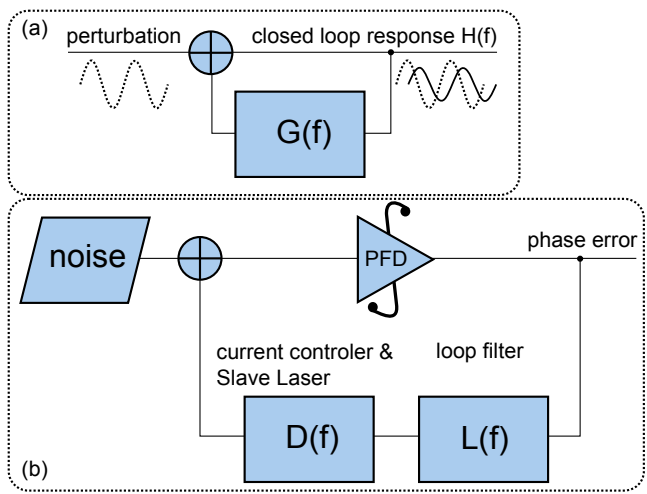}
\caption{(a) Schematic representation of closed loop response measurement. Perturbing signal is summed with loop response. Output is observed just after the summator. (b) OPLL equivalent schematic showing loop elements as transfer functions.  }
\label{fig:control_theory}       
\end{figure}

In this section we present a model of the OFLL used to select optimal loop parameters. We assume an OFLL as sketched in Fig.~\ref{fig:control_theory}. 
The transfer function of the open loop $G(f)$ is the product of constituents transfer functions 
$G(f) = {A\: D(f) L(f)}/{2 \pi i f},$ 
where $A$ is the overall gain constant, $D(f)$ and $L(f)$ are laser and loop filter transfer functions respectively and the integration due to the frequency-phase conversion is explicitly included.

In optimization procedure we consider the closed loop which transfer function is given by $H(f)=1/(1+G(f))$. In turn the eigenfrequencies of the closed loop are given by the roots of the $H(f)$ denominator. Their location on the complex plane determines the time response as well as the stability of the loop. The necessary condition for non-oscillatory behavior is that all roots lay in the negative real half plane. The root of interest for optimization purposes is the one with the highest real part as that is approximately equal inverse of the loop response time.

To maintain generality and simplicity we assume laser response $D(f)$ to be flat with dominant loop delay $\Delta \tau$,
$
D(f) = \exp\left( - i 2 \pi f \Delta \tau \right).
$

We first consider a purely proportional loop filter i.e. $L(f)=1$ with proportional gain constant included in $A$. 
The optimal gain in this case is found to be $A_{P,\mathrm{opt}} = 1/({e\Delta \tau})$. 

For the sake of 
further optimization considerations,
we also note that when the gain is increased 
up to the stability margin, oscillations develop at a frequency $f_{P,u} = {1}/{(4 \Delta \tau )}$ depending only on the loop delay $\Delta \tau$. 

The dependence originates from the stability condition requiring the gain to be below unity at $180\degree$ phase shift. Upon the condition violation the overall integration factor is responsible for one $90\degree$ and the loop delay $\exp\left(-i 2\pi f \Delta \tau\right)$ for the other introducing the $\frac{1}{4}$ factor.

An addition of the integral term to the loop filter compensates, otherwise steadily accumulating, total phase error. Therefore, it can be kept within the finite operating range of the phase detector.

In this case the loop filter $L(f)$ is characterized by a large DC gain $\kappa$ and filter zero frequency $f_z$ having a unity proportional gain at high frequencies:
\begin{equation}
\centering
L(f) = \kappa\frac{1+{i f}/{f_z }}{1+{i f \kappa }/{ f_z }}.
\label{eq:Lf}
\end{equation}
At $f_z$ the gain contributions from the proportional and integral terms are equal.

The presented model can be used to predict open loop time response and optimize $f_z$ and $A$ for a given $\Delta \tau$ aiming at quickest, non-oscillatory response. For large DC gains ($\kappa\gg 1$) the optimal parameters do not depend on $\kappa$. They are found to be very close to $A = 1/(2 \Delta \tau)$ and $f_z = {1}/({10\pi\Delta \tau})$.

The model definitions and optimization calculations can be found in a Mathematica worksheet in attached supplementary materials, allowing easy extension and adaptation to parameters measured in particular experiment.

\section{Setting up and optimization}
An experimental procedure to quickly set up the OFLL relies only on the measurements in the time domain thus eliminating the need for specialized devices such as network analyzers. The procedure aims at establishing the loop delay $\Delta \tau$ and its total gain. During the measurement, a care should be taken to ensure a wide PFD phase detection range by setting a high N divisor value. This prevents loop oscillations originating from PFD overflow. Indications of these can be observed as discontinuities in the PFD output signal.

First, a proportional amplifier is used to close the loop with a gain $P$ so chosen as to avoid oscillations.
The proportional constant $P$ should be subsequently increased until the loop develops oscillations. When these arise the total gain of the loop equals 1 and their frequency corresponds to the loop delay $f_{P,u} = {1}/{(4 \Delta \tau )}$. This allows to immediately use the optimal PI parameters by merely correcting for the gain $P_P$ used to arrive at oscillations. The derived model parameters correspond to the proportional gain $P_{PI} = A P_P / (2 \pi f_{P,u})$ and integral gain $I = 2 \pi P_{PI} f_z$. These equations can be combined with derived optimal parameter values to obtain a formula for the optimal PI gains in terms of used $P_P$ and measured $\Delta \tau$:
$P_{PI} = P_P / \pi$, $I = P_{PI}/(5 \Delta \tau)$.

\section {Tailoring the model}

\begin{figure}
 \includegraphics[width=\columnwidth]{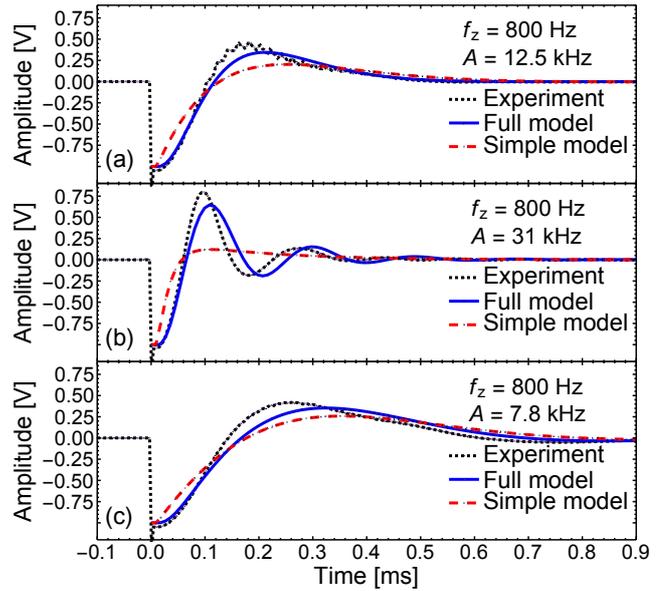}
\caption{ Experimental and theoretical closed loop time response for unit step perturbation. Optimal response (a) occurs for parameters $f_z = 800~\textrm{Hz}$, $A = 12.5~\textrm{kHz}$. Upon increasing $f_z$ damping period is prolonged (b). If the $f_z$ is lowered to much to suppress oscillatory behavior one may obtain slow sub-optimal response c). 
}
\label{fig:models_comp}       
\end{figure}

The simplicity of the model presented in Sec.~\ref{sec:smodel} provides its usability in a generic case. This, however, comes at the cost of omitting details in the loop elements description. It is quite straightforward to adjust the model if the details of the involved transfer functions are known. As depicted in Fig. \ref{fig:models_comp}, we have compared the experimental data with simple model predictions and these made after accounting for the measured laser current controller response. In our case it comprised both $10~\mu\textrm{s}$ delay and a double pole at $f=12.8~\textrm{kHz}$. In Fig. \ref{fig:models_comp}(b) it is evident that the simple model does not capture details of the loop time response. Here, according to the simple model, an increase in the loop gain relative to the situation depicted in Fig. \ref{fig:models_comp}(a) shall only improve the response time while the extended model properly captures the appearance of the unwanted oscillatory behavior.

The most prominent alterations to the model can be made by adjusting the laser response $D(f)$. To construct the corresponding transfer function, its complex poles and zeros may be located. This can be achieved by subsequently applying harmonic perturbations at different frequencies to the loop and recording the phase shift and amplification of the response. The sole $D(f)$ can be established either directly from the measurements of microwave phase or inferred from the loop response. The former requires additional apparatus depicted in Fig. \ref{fig:setup}(a) while the latter assumes $D(f)$ to be the only unknown transfer function in the model. Validity of any alteration of the model can be verified to some degree by measuring the loop time response.  

Accounting for the additional phase shifts in our setup we find the optimal gain and PI zero parameter values lower by a factor of $4$ compared to the predictions of the simple model.

\begin{align}
\centering
A = {1}/{(8 \Delta \tau)}, \qquad 
f_z = {1}/({40\pi\Delta \tau}).
\end{align}
Therefore, we advise to use reduced gain $A$ and zero frequency $f_z$. 

\section{Performance}

Here we discuss the system performance after application of the optimization procedures.
In our configuration the closed loop time response was found to be around $100\:\textrm{$\mu$s}$ as depicted in Fig. \ref{fig:models_comp}(a). 

\begin{figure}
 \includegraphics[width=\columnwidth]{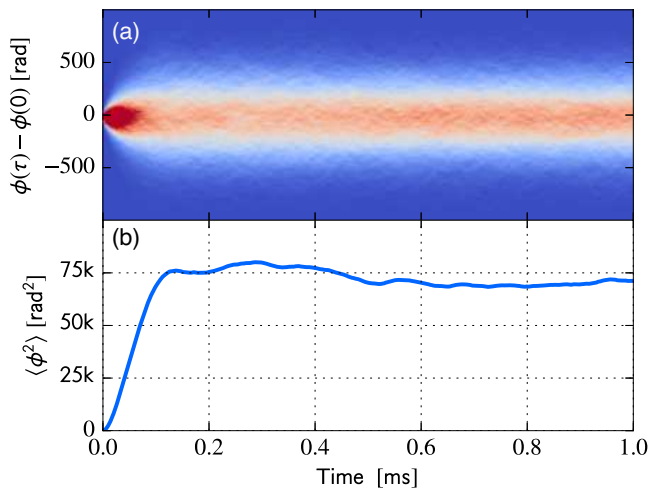}
\caption{Phase evolution of the system: (a) ensemble of lasers beat note phase deviations $\phi(t)$ with matched zero level and (b) phase variance time evolution showing. We infer the loop response time $\approx 100\:\mu\textrm{s}$ and the typical phase deviation $\approx 270\:\textrm{rad}$.}
\label{fig:phase}       
\end{figure}

\begin{figure}
 \includegraphics[width=\columnwidth]{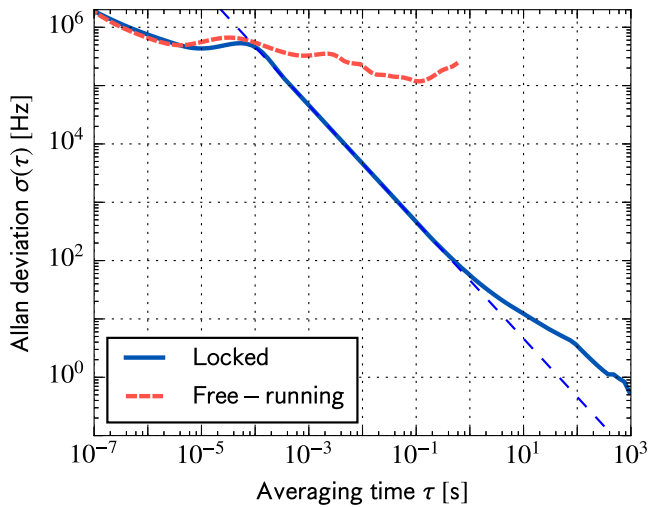}
\caption{Overlapping Allan deviation for locked and free-running SL showing the effect of the feedback loop for averaging times over $100$~$\mu\mathrm{s}$. We observe excellent behavior at long averaging times corresponding to a strong noise suppression at low frequencies. Dashed blue line corresponds to the $t^{-1}$ trend due to phase white noise.}
\label{fig:allan}       
\end{figure}

Performance and short-term stability is well characterized by the phase evolution of the system. As depicted in Fig.~\ref{fig:setup}(a) the phase $\phi (t)$ has been measured using the apparatus consisting of an in-phase quadrature mixer (I/Q mixer) ADL5380 fed a reference signal (LO’) from HMC833 high frequency generator. The mixer measures the in-phase $I_{iq}$ and $90 \degree$ shifted $Q_{iq}$ components of the RF signal enabling the retrieval of the phase $\phi (t) = \mathrm{arctan}\left(Q_{iq}/I_{iq}\right)$ relative to the LO’ . By collecting many 1 ms long records of $\Phi (t)$ and shifting the zero level of each by a constant, we obtained the record of $\phi(\tau)-\phi(0)$ and depicted it in Fig.~\ref{fig:phase}. As $\left<\phi(\tau)\right> = 0$, we obtain the phase variance as $\left<\phi(\tau)^2\right>$. 

The phase variance $\left<(\phi(t+\tau)-\phi(t))^2\right>$ reaches a constant value of $75\times10^3 \:\textrm{rad}^2$ after the settling time of $100\:\textrm{$\mu$s}$ confirming the correct functioning of the OFLL. The N divider was set to be $N=12800$. This corresponds to the PFD range of around $4 \times 10^4\:\textrm{rad}$. Thus the PFD range was kept far above the average phase deviation, allowing simple adjustments of the loop gain by altering $N$ values. 

We analyzed system capabilities for a broad range of offset frequencies. In particular, Fig. 5 presents the dependence of the PFD signal output on laser detuning for a set of reference frequencies. We find that the signal is suitable for locking for frequency offsets ranging from 4 to 15 GHz.

Finally, we characterize the long-term stability of the system. A convenient tool is the overlapping Allan variance, which is calculated using two separate phase measurements. To obtain the result for short averaging times, we acquired a single 1.4-s-long record of phase using the I/Q mixer. For longer averaging times we continuously measured PFD output signal for 2800 s and calculated the corresponding phase. Figure \ref{fig:allan} depicts the result for overlapping Allan deviation for locked and unlocked (free-running) laser. Long averaging times region is dominated by the phase white noise as inferred from the observed $\tau^{-1}$ power-law dependence. In this region, starting from $\tau\approx100\:\mu\mathrm{s}$ we clearly see the effect of the OFLL. In particular, the long-term stability of our system is guaranteed by an intrinsically environmentally-insensitive frequency detection, resulting in $<$ 0.5 Hz frequency deviation after 1000 s averaging time. At small averaging time the $\tau^{-1}$ trend discontinues due to a finite laser linewidth.

\section{Conclusions}
In conclusion, we have presented a laser difference stabilization technique extending the optical phase locked loop methods to the regime of broad-line lasers. 
We have discussed the OFLL operation, constructing a simple model which enabled us to present a simple method of OFLL optimization, relying merely on a straightforward time domain measurement. Furthermore, the loop diagnostic methods have been presented.   

Our setup provides an excellent long-term frequency stability, yet providing a broad lock set point frequency range (4--15~GHz) and extreme capture range, promising a variety of applications in quantum optics and cold atomic physics. If smaller offset frequencies are required similar PFD chip suitable for smaller frequencies can be used.
Furthermore, a phase variance measurement on the PFD output provides a simple method to eliminate an unwanted locking at zero frequency offset thus extending the effective capture range to nearly 30~GHz. 

Exploiting an integrated phase frequency detector ADF41020 on an ordinary PCB with merely one microwave track, our design remains simple and readily compatible with generic laser current controllers. Digitally controlled ADF41020 frequency divisors and lock setpoint (e.g. by DDS AD9959) allow for simple regulation. The ADF41020 can independently output the divided measured frequency, which enables construction of fully automated, self-diagnosing systems with standard counters.

\section*{Acknowledgments}
We acknowledge contributions of M. Piasecki at the early stages of the project, input of J. Iwaszkiewicz to the development of microcontroller-based test setup as well as support of K. Banaszek and R. Łapkiewicz and careful proofreading of the manuscript by M. D\k{a}browski.
This research was funded by the National Science Center (Poland) Grants No. 2015/17/D/ST2/03471 and 2015/16/S/ST2/00424 and by the
Polish Ministry of Science and Higher Educations "Diamentowy Grant" Project No. DI2013 011943.

\bibliographystyle{spphys}       
\bibliography{ofll}   
\end{document}